\documentclass{aastex}
\usepackage{spr-astr-addons}
\usepackage{url}
\usepackage{verbatim}

\RequirePackage{color}

\newcommand{\emaila}

\begin{document}

\title{Noether gauge symmetry for $f(R)$ gravity in Palatini formalism}

\shorttitle{Noether symmetries of Bianchi type I spacetime}
\shortauthors{Y. Kucukakca and U. Camci}

\author{Y. Kucukakca and U. Camci}

\affil{Department of Physics, Faculty of Science, Akdeniz
University,
\\ \quad 07058 Antalya, Turkey \\
\emaila{ykucukakca@gmail.com and ucamci@akdeniz.edu.tr}}


\begin{abstract}
In this study, we consider a flat Friedmann-Robertson-Walker (FRW)
universe in the context of Palatini $f(R)$ theory of gravity. Using
the dynamical equivalence between $f(R)$ gravity and scalar-tensor
theories, we construct a point Lagrangian in the flat FRW spacetime.
Applying {\em Noether gauge symmetry approach} for this $f(R)$
Lagrangian we find out the form of $f(R)$ and the exact solution for
cosmic scale factor. It is shown that the resulting form of $f(R)$
yield a power-law expansion for the scale factor of the universe.
\end{abstract}


\today


\section{Introduction}

It is well-known that the Einstein-Hilbert action produces general
relativity (GR) which leads mostly to observational success.
Furthermore the modified gravity theories issued by more general
gravitational actions could explain the observational facts.
Observations of the supernovae Type Ia \cite{riess99,
perlmutter99} and the cosmic microwave background radiation
\cite{netterfield02} indicate that current expansion of the
universe is accelerating, on the contrary to
Friedmann-Robertson-Walker (FRW) solution of GR. To explain such
an accelerated expansion, in the framework of GR, many authors
introduced mysterious cosmic fluid, the so-called \emph{dark
energy} which can be described by the cosmological constant
\cite{copeland06,durrer08}.\\ On the other hand, to get a solution
of this problem, some modifications in GR theory have been
proposed. The $f(R)$ gravity theory is a particular class of
modified gravity theories \cite{fr,nojiri07,felice10}. The
simplest form of this theory can be constructed by replacing the
Ricci scalar $R$ with an arbitrary function $f(R)$ in the
Einstein-Hilbert Lagrangian (EHL). In recent literature some
different forms of $f(R)$ gravity have been proposed, and
discussed in different contexts. For example, it has been shown
that early-time inflation and current cosmic acceleration may take
place by adding negative and positive powers of curvature into the
EHL \cite{nojiri03}. Carroll et al. \cite{Carroll04} have also
proposed that a small correction to the EHL by adding an inverse
term of $R$ would lead to cosmic speed-up which originates from
purely gravitational effects. Similar modifications of GR have
also been proposed to drive inflation \cite{staro80}. On the other
hand, it is worth noting that the main deficiency of such theories
is that they are seriously constrained by solar system
tests\cite{olmo05,chiba07}. A number of viable $f(R)$ models that
can satisfy both cosmological and local gravity constraints have
been proposed in Refs. \cite{Amendola07,star07,cognola08}.

There are two kinds of Noether symmetry approach for cosmological
studies in the literature : The first one is the so-called Noether
symmetry approach in which the Lie derivative of a given
Lagrangian vanishes,
\cite{capo93,ritis,demianski92,sanyal01,camci07,souza} and the
second one is the so-called Noether gauge symmetry (NGS) approach
\cite{jamil,hussain11}. The latter is a generalization of the
former Noether symmetry approach in the sense that Noether
symmetry equation includes a gauge term. Taking into account a
gauge term in Noether symmetry equation gives a more general
definition of the Noether symmetry: this is the so-called Noether
gauge symmetry (NGS) approach. Thus one may expect extra (more
than one) symmetry generators from this definition.

We note that the Noether symmetry approach without gauge term
allows one to choose the potential dynamically in the
scalar-tensor gravity theory \cite{sanyal03}, and very recently,
the explicit form of the function $f(R)$ \cite{Cap08,Vakili08}.
For this form of $f(R)$, cosmological solutions of FRW metric can
describe the accelerated period of the universe. Applying this
Noether symmetry approach, the spherically symmetric solutions in
$f(R)$ theories of gravity have been also found in \cite{Capp07}.
Using the definition of NGS, some authors \cite{jamil,hussain11}
have discussed $f(R)$ and $f(R)$-Tachyon model in the metric
formalism, where they conjectured that the application of NGS to
generic $f(R)$ Lagrangian results in zero gauge function. The
Palatini formalism has not been considered by using the NGS
approach in the literature yet. In a recent work Roshan and Shojai
\cite{fat} studied Palatini $f(R)$ cosmology in flat FRW
spacetimes following Noether symmetry approach without gauge term
for the matter dominated universe. They found out the form of
$f(R)$ as power-law and exact solutions for cosmic scale factor.
In this study we have generalized their result considering NGS
approach within the scope of the Palatini $f(R)$ gravity.

This paper is organized as follows. In the following section, the
Palatini formalism is briefly reviewed. In section \ref{ngs}, we
discuss the Noether symmetry approach with/without a gauge term
for the Palatini $f(R)$ gravity. In section \ref{solution}, we
search the cosmological solutions by using the obtained forms of
$f(R)$. Finally, in Section \ref{conc}, we conclude with a brief
summary.

\section{The $f(\mathcal{R})$ Gravity and The Field Equations } \label{b1}

In  four dimensions, the action of the Palatini $f(\mathcal{R})$
gravity theory with matter is written as
\begin{eqnarray}
\mathcal{A}  =  \int{d^4 x \sqrt{-g} \left[ f(\mathcal{R}) + 2\kappa
L_{m}(g_{a b},  \psi) \right]} . \label{action}
\end{eqnarray}
Here $f(\mathcal{R})$ is a differentiable function of the Ricci
scalar $\mathcal{R} = g^{a b}\mathcal{R}_{a b}(\Gamma)$,
$\mathcal{R}_{a b}(\Gamma)$ is the Ricci tensor of any independent
torsionless connection $\Gamma$ independent of $g_{ab}$ and $L_{m}$
is the matter Lagrangian and $\psi$ represents collectively the
matter fields. The matter Lagrangian is chosen as $L_{m} =
-\rho_{m0} a^{-3}$ for matter dominated cosmology. In general there
are two approach in order to derive the dynamical field equations of
motion in $f(R)$ gravity. The first one is Palatini approach
\cite{pala,olmo05b,fay07,bao07} in which the metric and connection
are considered as independent quantities, and the action is varied
with respect to both of them. The second approach is the metric
formalism in which action is varied with respect to metric tensor.
The field equations in the metric formalism are fourth-order
differential equations, while for Palatini formalism they are
second-order. If $f(R)$ is linear in $R$, the two approaches lead to
the same equation. In this study we will use the Palatini formalism.
Variation of the action (\ref{action}) with respect to metric tensor
yields following field equations \cite{Brax08}
\begin{equation}
f_{\mathcal{R}} \mathcal{R}_{a b}-\frac{1}{2} f g_{a b} = \kappa
T_{a b}, \label{f-eq1}
\end{equation}
and the variation of the action (\ref{action}) with respect to the
connection gives
\begin{equation}
\nabla_{c}\left( \sqrt{-g}f_{\mathcal{R}} g^{a b}\right) = 0,
\label{f-eq2}
\end{equation}
where $f_{\mathcal{R}} = df/d \mathcal{R}$ and $T_{a b}$ is the
usual stress-energy tensor of the matter. Also, the trace of Eq.
(\ref{f-eq1}) is
\begin{equation}
f_{\mathcal{R}} \mathcal{R}-2f = \kappa T. \label{trace}
\end{equation}
The $f(\mathcal{R})$ gravity theory in Palatini formalism is
equivalent to $\omega_{BD} = -3/2$ a Brans-Dicke (BD) theory. In
order to construct a canonically effective point-like Lagrangian,
we have to use the dynamical equivalence between Palatini
$f\mathcal({R})$ formalism and the BD theory of gravity
\cite{st06,cap11}. Therefore the action (\ref{action}) can be
written as follows
\begin{eqnarray}
\mathcal{A} = \int{d^4 x \sqrt{-g}(\phi R + \frac{3}{2\phi}g^{a b}
\partial_{a} \phi \partial_{b} \phi - U(\phi) + 2 \kappa L_m
) }, \label{action-1}
\end{eqnarray}
where $\phi=f_{\mathcal{R}}$, $U(\phi)=\phi \chi(\phi) -
f(\chi(\phi))$, $\mathcal{R}=\chi(\phi)$ and $R$ is a curvature
scalar constructed from the Levi-Civita connection of metric tensor.
We note that this is the action of the BD theory with the BD
parameter $\omega_{BD} = -3/2$. It was also shown that in the metric
formalism the $f(R)$ gravity is equivalent to the BD theory with
$\omega_{BD} = 0$ \cite{fr}.

\par
In this study, we consider a matter-dominated model in a spatially
flat FRW universe with signature ($+,+,+,-$). Using the action
(\ref{action-1}) with the scalar curvature for FRW metric in the
form $R= 6\left( \ddot{a}/a + \dot{a}^2 / a^2 \right)$ and taking
$L_m = -\rho_{m0} a^{-3}$, a point-like Lagrangian takes such a
form \cite{fat}
\begin{eqnarray}
 L =   6 a^2\dot{a}\dot{\phi} + 6\phi a \dot{a}^2 +
\frac{3a^3}{2\phi} \dot{\phi}^2 + a^3 U(\phi) + 2\kappa \rho_{m0},
\label{lag}
\end{eqnarray}
where $a(t)$ is the cosmic scale factor, $\rho_{m0}$ is an
integration constant associated with matter content and the dot
indicates the derivation with respect to cosmic time $t$. It is
found that the \emph{energy function} $E_{L}$ associated with the
Lagrangian (\ref{lag}) vanishes, i.e.
\begin{eqnarray}
E_{L} = \frac{\dot{a}^2}{a^2} + \frac{\dot{a}\dot{\phi}}{a\phi} +
\frac{1\dot{\phi}^2}{4\phi^2} - \frac{U}{6 \phi} - \frac{\kappa
\rho_{m0}}{3 a^3\phi} =0, \label{E-L}
\end{eqnarray}
which is known as the modified Friedmann equation. The equations
of motion can be obtained by varying the Lagrangian (\ref{lag})
with respect to $a$ and $\phi$, respectively, as follows
\begin{eqnarray}
\frac{2\ddot{a}}{a} + \frac{\dot{a}^2}{a^2} +
\frac{\ddot{\phi}}{\phi} + \frac{2\dot{a}\dot{\phi}}{a\phi} -
\frac{3\dot{\phi}^2}{4\phi^2} - \frac{U}{2 \phi} =0, \label{de-1}
\end{eqnarray}
\begin{eqnarray}
\frac{\ddot{a}}{a} + \frac{\dot{a}^2}{a^2} +
\frac{\ddot{\phi}}{2\phi} + \frac{3\dot{a}\dot{\phi}}{2a\phi} -
\frac{\dot{\phi}^2}{4\phi^2} - \frac{U'}{6} =0. \label{de-2}
\end{eqnarray}

\section{The Noether symmetry approach} \label{ngs}

For most of studies in the context of both $f(R)$ gravity and the
scalar-tensor theory, it has been used definition of Noether
symmetry without a gauge term \cite{capo93,kamilya,Vakili08} ,
\cite{fat}. Recently, some work on NGS approach for FRW metric has
been appeared \cite{jamil,hussain11}. Noether gauge symmetry is
defined as follows. Let us consider a vector field \cite{ibrahim}
\begin{equation}\label{vecf}
{\bf X} = \tau \frac{\partial}{\partial t}+\alpha
\frac{\partial}{\partial a} + \beta \frac{\partial}{\partial \phi}
\end{equation}
where $\tau, \alpha$ and $\beta$ are depend on $t,a,$ and $\phi$.
Here, $t$ is the independent variable, $a(t)$ and $\phi(t)$ are
the dependent variables. The first prolongation of the above
vector field is given by
\begin{equation}\label{prol}
{\bf X^{[1]}} = {\bf X}+\alpha_{t} \frac{\partial}{\partial
\dot{a}} + \beta_{t} \frac{\partial}{\partial \dot{\phi}} ,
\end{equation}
in which
\begin{equation}\label{def}
{\alpha_{t}} = D_{t}\alpha-\dot{a}D_{t}\tau, \quad {\beta_{t}} =
D_{t}\beta-\dot{\phi}D_{t}\tau,
\end{equation}
where $D_{t}$ is the operator of total differentiation with
respect to $t$
\begin{equation}\label{totd}
{D_{t}} =  \frac{\partial}{\partial t} + \dot{a}
\frac{\partial}{\partial a} + \dot{\phi} \frac{\partial}{\partial
\phi}.
\end{equation}
The vector field $\bf X$ is a NGS of a Lagrangian
$L(t,a,\phi,\dot{a},\dot{\phi})$ if there exists a gauge function,
$B(t,a,\phi)$, such that
\begin{equation}\label{Noether}
{\bf X}^{[1]}L + LD_{t}(\tau) = D_{t}B .
\end{equation}
For Noether symmetry without gauge term (i.e. $B=0$) it is
required that $\tau=0$, and thus the above equation reduces to the
form $\pounds_{\bf X^{[1]}} L = 0$ which is the existence
condition for Noether symmetry without gauge term. The
significance of NGS is clear from the following theorem
\cite{noet}.

\emph{Theorem}: If ${\bf X}$ is the Noether symmetry corresponding
to the Lagrangian $L(t,a,\phi,\dot{a},\dot{\phi})$, then
\begin{equation}\label{FrstI}
{I} = \tau L + \left(\alpha-\tau \dot{a}\right) \frac{\partial
L}{\partial \dot{a}} + (\beta-\tau \dot{\phi}) \frac{\partial
L}{\partial \dot{\phi}} - B
\end{equation}
is a first integral or a conserved quantity associated with ${\bf
X}$. Now we seek the condition in order that the Lagrangian
(\ref{lag}) would admit NGS.

For the flat FRW metric, the NGS condition (\ref{Noether}) yields
the following set of equations
\begin{equation}
\tau_{a} = \tau_{\phi} = 0, \label{neq6}
\end{equation}
\begin{equation}
 6 a (2\phi \alpha_{t} + a \beta_{t}) - B_{a}= 0,
\label{neq1}
\end{equation}
\begin{equation}
\frac{3 a^2}{\phi} (2\phi \alpha_{t} + a \beta_{t}) - B_{\phi}= 0,
\label{neq2}
\end{equation}
\begin{equation}
2 \alpha + a  \left( \alpha_{a} + \beta_{\phi} - \tau_{t}\right) +
2\phi \alpha_{\phi} + \frac{a^2}{2\phi} \beta_{a} = 0,
\label{neq3}
\end{equation}
\begin{equation}
\phi \alpha + a \beta + a \phi \left( 2 \alpha_{a} -
\tau_{t}\right) + a^2 \beta_{a}= 0, \label{neq4}
\end{equation}
\begin{equation}
3 \alpha -\frac{a}{\phi} \beta  + a \left(2\beta_{\phi} -
\tau_{t}\right) + 4\phi \alpha_{\phi} = 0, \label{neq5}
\end{equation}
\begin{equation}
a^2 (3\alpha + a \tau_{t})U + a^3 U'\beta  +  2\kappa
\rho_{m0}\tau_{t} - B_{t}= 0. \label{neq7}
\end{equation}
After some algebraic calculations,  the solutions of the above set
of differential equations (\ref{neq1})-(\ref{neq7}) for $\alpha,
\beta$, $ \tau, B$ and the potential $U(\phi)$ are obtained as
follows:
\begin{eqnarray}
\alpha = c_{1} a + c_{2} a (a^2\phi)^{-\frac{\ell+1}{2\ell+1}},
\label{alpha}
\end{eqnarray}
\begin{eqnarray}
\beta = -(2\ell-1) \phi\left[\frac{3 c_{1} }{\ell-2} - c_{2}
(a^2\phi)^{-\frac{\ell+1}{2\ell+1}}\right] , \label{beta}
\end{eqnarray}
\begin{eqnarray}
\tau = -\frac{3c_{1}(\ell+1)}{\ell-2} t + c_{3},\, B =
-\frac{6c_{1}\kappa \rho_{m0}(\ell+1)}{\ell-2} t +
c_{4}\label{vec2}
\end{eqnarray}
\begin{eqnarray}
U(\phi) = \lambda \phi^{-\frac{3}{2\ell-1}} , \label{pot}
\end{eqnarray}
where $ c_i, \ell$ and $\lambda$ are constants and  $\ell\neq
-1/2, 2$. Since $B=0$ and $\tau=0$ for Noether symmetry without
gauge term, the parameters $c_1, c_3$ and $c_4$ should vanish.
This case has been studied in Ref.\cite{fat}. Taking $\phi =
\varphi^2$ in the Lagrangian (\ref{lag}) and doing the above
calculations again, the obtained form of $\alpha$ and $\beta$
generalizes the ones found in Ref.\cite{fat} (which are
represented as $A$ and $B$ in this reference).

It is seen from (\ref{alpha})-(\ref{vec2}) that the Lagrangian
(\ref{lag}) admits \emph{three} Noether symmetry generators
\begin{equation}
{\bf X}_1 = \frac{\partial}{\partial t}, \label{gen-1}
\end{equation}
\begin{equation}
{\bf X}_2 = (a^2 \phi)^{-\frac{(\ell+1)}{(2\ell+1)}}\left(a
\frac{\partial}{\partial a} + (2\ell-1) \phi
\frac{\partial}{\partial \phi}\right), \label{gen-2}
\end{equation}
\begin{equation}
{\bf X}_3 = -\frac{3(\ell+1)t}{\ell-2} \frac{\partial}{\partial t} +
a \frac{\partial}{\partial a} - \frac{3 (2\ell-1)\phi}{\ell-2}
\frac{\partial}{\partial \phi} .\label{gen-3}
\end{equation}
The corresponding Lie algebra has the following non-vanishing
commutators:
\begin{eqnarray} \label{commu1}
& & [{\bf X}_{1},{\bf X}_{3}] = -\frac{3(\ell+1)}{\ell-2} {\bf
X}_{1}, \\ & &  \label{commu2} [{\bf X}_{2},{\bf X}_{3}] =
-\frac{(\ell+1)(4\ell+1)}{(\ell-2)(2\ell+1)} {\bf X}_{2}.
\end{eqnarray}
The first integrals associated with ${\bf X_{i}}$ are
\begin{eqnarray} \label{frstI-1}
I_{1} = 6a^2 \dot{a}\dot{\phi} + 6a \phi \dot{a}^2 +
\frac{3a^3}{2\phi} \dot{\phi}^2 - a^3 U - 2\kappa \rho_{m0},
\end{eqnarray}
\begin{eqnarray} \label{frstI-2}
I_{2} =  3(2\ell+1) a (a^2 \phi)^{-\frac{(\ell+1)}{(2\ell+1)}} \
\frac{d}{dt} (a^2 \phi),
\end{eqnarray}
\begin{eqnarray} \label{frstI-3}
I_{3} = \frac{3(\ell+1)}{\ell-2} t I_{1} - \frac{3(4\ell+1)a
}{\ell-2} \ \frac{d(a^2 \phi)}{dt} \nonumber
\\ + \frac{6(\ell+1) \kappa
\rho_{m0}}{\ell-2}t.
\end{eqnarray}
Here the constant parameter $c_{4}$ is assumed to be zero in the
gauge function $B$. We note that the first integral
(\ref{frstI-1}) is related with the \emph{energy function}
(\ref{E-L}), so that the first integral $I_{1}$ vanishes.

\section{The Cosmological Solution} \label{solution}

As an inverse problem of finding $f(\mathcal{R})$ Lagrangian, it
is only required to give $U(\varphi)$. Using the algebraic
relation
\begin{equation}
U(\phi)=\phi \chi(\phi) - f(\chi(\phi)), \label{pot-bag}
\end{equation}
it is possible to solve $f(\mathcal{R})$, where
$\mathcal{R}=\chi(\phi)$ and $\phi = f_{\mathcal{R}}$. Putting the
potential (\ref{pot}) into Eq. (\ref{pot-bag}) yields
\begin{equation}
f_{\mathcal{R}}\mathcal{R}-f = \lambda
(f_{\mathcal{R}})^{-\frac{3}{2\ell-1}}.\label{defe}
\end{equation}
Thus, it is straightforward to get the following two solutions
from Eq. (\ref{defe}) for $f(\mathcal{R})$
\begin{equation}
f(\mathcal{R}) = p (\ell) \mathcal{R}^{\frac{3}{2(\ell+1)}}
,\label{form1}
\end{equation}
\begin{equation}
f(\mathcal{R}) = R_{0} \mathcal{R} - \lambda
R_{0}^{-\frac{3}{2\ell-1}},\label{form2}
\end{equation}
where  $p(\ell)= 2(\ell +1) \left[ \frac{\lambda 3^{-3/(2 \ell
-1)}}{1- 2\ell} \right]^{\frac{2 \ell -1}{2(\ell+1)}}$, $R_{0}$ is
a constant and $\ell\neq -1, 1/2$.

In the following subsections, considering $f(\mathcal{R})$ given
by (\ref{form1}) and (\ref{form2}), we will search the exact
solution for cosmic scale factor. For $\alpha$ and $\beta$ given
by (\ref{alpha})-(\ref{beta}) and Eq. (\ref{form1}), it is
appeared the singularities for the values of $\ell =-1/2$ and
$\ell = -1$. For these values of $\ell$, after solving
Eq.(\ref{defe}) one has the same form with (\ref{form2}) for
$f(\mathcal{R})$.
\par

\subsection{Case (i).  $f(\mathcal{R}) = p(\ell)
\mathcal{R}^{\frac{3}{2(\ell+1)}}$ } \label{case-i}

Using the trace relation (\ref{trace}) and $\phi=f_{\mathcal{R}}$ we
obtain
\begin{equation}
\phi = r(\ell) a^{2\ell-1}, \label{trace1}
\end{equation}
where $r(\ell)=\frac{3
p(\ell)}{2(\ell+1)}[\frac{2\kappa\rho_{m0}(\ell+1)}{p(\ell)(4\ell+1)}]^{\frac{1-2\ell}{3}}$,
$\ell \neq -1/4$. Hence, the Noether first integrals (\ref{frstI-2})
and (\ref{frstI-3}) can be written as
\begin{eqnarray} \label{frstI-22}
I_{2} = 3(2\ell+1)^2 r(\ell)^{\frac{\ell}{2\ell+1}} a^{\ell}
\dot{a},
\end{eqnarray}
\begin{eqnarray} \label{frstI-33}
I_{3} = -\frac{3(4\ell+1)(2\ell+1) r(\ell)}{\ell-2} a^{(2\ell+1)}
\dot{a} \nonumber
\\ + \frac{6\kappa \rho_{m0}(\ell+1)}{\ell-2}t.
\end{eqnarray}
Now the Eq.(\ref{frstI-22}) can be used to find out the time
dependence of the cosmic scale factor as
\begin{eqnarray} \label{sol-1}
a(t) = \left[\frac{I_{2}(\ell+1)
r(\ell)^{-\frac{\ell}{2\ell+1}}}{3(2\ell+1)^2} t
 + t_{0}\right]^{\frac{1}{\ell +1}},
\end{eqnarray}
where $t_0$ is  a constant of integration. Considering $a(t)$ in
Eq.(\ref{frstI-33}), the following constraint equations are found
\begin{eqnarray}
I_{2}^2 (4\ell+1) r(\ell)^{1/(2\ell+1)} - 18
\kappa\rho_{m0}(2\ell+1)^3 =0 , \label{ceq-1}
\end{eqnarray}
\begin{eqnarray}
I_{2} t_{0} (4\ell+1) r(\ell)^{(\ell+1)/(2\ell+1)} +
(\ell-2)(2\ell+1)I_{3} = 0. \label{ceq-2}
\end{eqnarray}
It is explicitly seen here that if $t_{0}=0$, then the first
integral $I_3$ is equal to zero. Also, using Eqs.(\ref{E-L}),
(\ref{de-1}) and (\ref{de-2}), we have additional constraint
relations as
\begin{eqnarray}
\frac{I_2^2 (2\ell-1) r(\ell)^{1/(2\ell+1)}}{18(2\ell+1)^3} +
\lambda
 r(\ell)^{-3/(2\ell-1)} =0 , \label{ceq-3}
\end{eqnarray}
\begin{eqnarray}
\frac{I_{2}^2  r(\ell)^{1/(2\ell+1)}}{6 (2\ell +1)^2 } - \left[
\lambda \, r(\ell)^{-3/(2\ell-1)} + 2 \kappa\rho_{m0} \right] = 0.
\label{ceq-4}
\end{eqnarray}
Thus, the constraint Eqs. (\ref{ceq-1}) and (\ref{ceq-3}) give the
following simple relation
\begin{eqnarray}
\kappa \rho_{m0} = - \frac{\lambda (4\ell+1)}{2\ell-1}
r(\ell)^{-3/(2\ell-1)}. \label{ceq-13}
\end{eqnarray}

The deceleration parameter, which is an important quantity in
cosmology, is defined by $q =-a \ddot{a}/\dot{a}^2$ , where the
positive sign of $q$ indicates the standard decelerating models
whereas the negative sign corresponds to accelerating models and
$q=0$ corresponds to expansion with constant velocity.  It takes
the following form in this model
\begin{eqnarray} \label{dec}
& &  q = \ell.
\end{eqnarray}
The effective equation of state parameter defined by
$w_{eff}=-1-\frac{2\dot{H}}{3H^2}=\frac{2q-1}{3}$
\cite{capo03,all04a,all04b} can be obtained as
\begin{eqnarray} \label{eos-1}
& &  w_{eff} =  \frac{2\ell-1}{3},
\end{eqnarray}
where $H$ is Hubble parameter. Astrophysical data indicate that $w$
lies in a very narrow strip close to $w = -1$. The case $w = -1$
corresponds to the cosmological constant. For $w<-1$ the phantom
phase is observed, and for $-1 < w < -1/3$ the phase is described by
quintessence. Thus, in the interval $-1 < \ell < 0$ we have
quintessence phase. If $-\infty < \ell < -1$, then the phantom phase
occurs, where the universe is both expanding and accelerating.
\par

\subsection{Case (ii). $f(\mathcal{R}) = R_{0} \mathcal{R} - \lambda
R_{0}^{-3/(2\ell-1) }$ } \label{case-ii}

\par
If $R_{0} = 1$, i.e. $\phi =1$ from the relation $\phi =
f_{\mathcal{R}} = R_0$, then the action is reduced to the
Einstein-Hilbert action with cosmological constant (
$f(\mathcal{R})=\mathcal{R}-\lambda$). We note here that Palatini
and metric formalism are coincide. In this case, the Noether first
integrals (\ref{frstI-2}) and (\ref{frstI-3}) can be written as
\begin{eqnarray} \label{frstII-0}
& &  I_{2} = 6(2\ell+1) a^{\frac{2\ell}{2\ell+1}} \dot{a},
\end{eqnarray}
\begin{eqnarray} \label{frstI-3-ii}
& &  I_{3} = -\frac{6(4\ell+1)}{\ell-2} a^2 \dot{a} +
\frac{6(\ell+1)\kappa \rho_{m0}}{\ell-2} t.
\end{eqnarray}
The modified Friedmann equation (\ref{E-L}) for this case reduces to
the form
\begin{eqnarray} \label{mfeq}
& &  \frac{\dot{a}^2}{a^2} - \frac{\lambda}{6} - \frac{ \kappa
\rho_{m0}}{3 a^3} = 0.
\end{eqnarray}
From Eq. (\ref{frstII-0}) the scale factor is solved as
\begin{eqnarray} \label{sol-2}
& &  a(t) = \left[\frac{ I_{2} (4\ell+1)} {6(2\ell+1)^2} t
 + t_{1}\right]^{\frac{2\ell+1}{4\ell+1}}.
\end{eqnarray}
Inserting the scale factor (\ref{sol-2}) into  Eq.
(\ref{frstI-3-ii}) and Eq. (\ref{mfeq}) one gets $\lambda =0$,
$\ell=1/2$ and as a constraint equations  $\kappa \rho_{m0}=
I_{2}^2/48$ and $t_{1}=I_{3}/I_{2}$. Therefore, for this case the
scale factor has the form
\begin{equation} \label{sol2-1}
a(t)=\left(\frac{I_{2}}{8} t + \frac{I_{3}}{I_{2}}\right)^{2/3}
\end{equation}
which is obtained from (\ref{sol-2}) when $\ell=1/2$ . Taking
$\ell=-1/4$ in Eq.(\ref{frstII-0}), we have $a(t)=\exp(\frac{I_{2}
t}{3})$ which gives $\rho_{m0}=0$ and $\lambda =  2 I_2^2 /3$.
Using these results for $\ell = -1/4$ in the Eq.
(\ref{frstI-3-ii}) one can find $I_{3}=0$.

\section{Concluding remarks}
\label{conc}

The Palatini approach consider the metric $g_{ab}$ and the
connection $\Gamma^{a}_{bc}$ as independent field variables, but
the spacetime metric $g_{ab}$ is only independent variable in
metric formalism. The Palatini formalism can be seen as containing
two independent metrics $g_{ab}$ and $h_{ab} = f_{\mathcal{R}}
g_{ab}$ rather that a metric and independent connection. In
Palatini $f(\mathcal{R})$ gravity the second metric $h_{ab}$
determine the geodesic structure with the connection
$\Gamma^{a}_{bc}$ which is the Levi-Civita connection of new
metric $h_{ab}$. In BD theory of gravity the second metric
$h_{ab}$ is related to the non-minimal coupling of the BD scalar,
i.e. $\phi = f_{\mathcal{R}}$. In Palatini approach,
$f(\mathcal{R})$ gravity given by the action (\ref{action}) is
equivalent to a special BD theory with a scalar field potential.
The action (\ref{action-1}) is clearly that of a BD theory with BD
parameter $w_{BD} = -3/2$ and a potential $U(\phi)$, which is
considered in this paper.  In the metric formalism, $f(R)$ gravity
is equivalent to the BD theory with $w_{BD} = 0$ (see the review
paper \cite{capo2011}). An $w_{BD} =0$ BD theory was originally
studied for the aim of getting Yukawa correction to the Newtonian
potential in the weak-field limit \cite{hanlon72}.

In this study, we examined the matter dominated flat FRW universe
by considering the Palatini $f(\mathcal{R})$ formalism and by
following the NGS approach which leads to explicit form for
$f(\mathcal{R})$. This approach is based on the search for Noether
gauge symmetries which allow one to find the form of
$f(\mathcal{R})$. We have obtained two type Noether symmetric
$f(\mathcal{R})$, (\ref{form1}) which yields case (i) and
(\ref{form2}) which gives case(ii), where $f(\mathcal{R})$
functions give rise to a power-law Lagrangian and EHL with
cosmological constant, respectively. In case (i), it is found the
power-law form of cosmic scale factor by (\ref{sol-1}). For case
(ii), the cosmic scale factor is obtained by (\ref{sol2-1}) for
$\ell = 1/2$ and de Sitter solution for $\ell = -1/4$. We have
presented the effective equation of state parameter for Palatini
$f(\mathcal{R})$ cosmology. In the first model, case(i), the
expansion of Universe is accelerating at the intervals $-1 < \ell
<0$, quintessence phase, and $-\infty < \ell < -1$, phantom phase.
Thus, this model can provide a natural gravitational alternative
for the dark energy without the necessity to introduce an exotic
fluid with a negative equation of state parameter. We note here
that while the gauge function turns out to be zero in
Refs.\cite{jamil,hussain11}, but our analysis shows that it
depends on the cosmic time.

\section*{Acknowledgements}

We are grateful to Prof. Naresh Dadhich for useful comments and
discussions. This work was supported by the Scientific Research
Projects Unit of Akdeniz University.

\end{document}